# Fractal hierarchy enables exponential scaling of topological boundary states

Limin Song[1†], Zhichan Hu[1†], Ziteng Wang[1], Domenico Bongiovanni[1,2], Liqin Tang[1], Daohong Song[1,3], Roberto Morandotti[2], Jingjun Xu[1], Hrvoje Buljan[1,4,*], and Zhigang Chen[1,3,*]

[1] *The MOE Key Laboratory of Weak-Light Nonlinear Photonics, TEDA Applied Physics Institute and School of Physics, Nankai University, Tianjin 300457, China*

[2] *INRS-EMT, 1650 Blvd. Lionel-Boulet, Varennes, Quebec J3X 1S2, Canada*

[3] *Collaborative Innovation Center of Extreme Optics, Shanxi University, Taiyuan, Shanxi 030006, China*

[4] *Department of Physics, Faculty of Science, University of Zagreb, Bijenička c. 32, 10000 Zagreb, Croatia*

[†] *These authors contributed equally to this work*

*hbuljan@phy.hr, zgchen@nankai.edu.cn

## Abstract

Exponential growth describes an extremely rapid process ubiquitous across mathematics and diverse physical, biological, and technological systems. Here, we introduce a class of fractal-inspired lattices that combine long-range periodic order with self-similar hierarchy, establishing a structural motif that enables exponential scaling of topological boundary states. We demonstrate this phenomenon in (i) a quasi-one-dimensional lattice chain constructed from Koch-curve unit cells and (ii) a two-dimensional periodic tiling lattice composed of Sierpiński-gasket unit cells. We show that, for suitable coupling parameters, both the number of topological boundary states $N_\ell$ and the number of topological minigaps $M_\ell$ grow exponentially with the fractal generation index $\ell$. We find that $N_\ell$ is an integer multiple of $M_\ell$, with the integer determined by the underlying symmetry. This hierarchical scaling law is captured by multi-topological-phase theory and confirmed experimentally in laser-written photonic lattices. Our results identify fractal hierarchy as a materials architecture principle for controlling boundary-state multiplicity, revealing an interplay between topology, self-similar geometry, and periodic order. More broadly, this work suggests a route to synthetic materials and integrated photonic platforms in which large numbers of robust boundary modes can be engineered within compact architectures.

**Introduction**

Topological phases of matter in periodic and aperiodic structures, including fractals and quasicrystals, have emerged as a fertile arena for discovering unconventional boundary phenomena. In periodic systems, integer-dimensional lattices endowed with chiral symmetry or discrete rotational ($C_n$) symmetry are described by a well-defined band theory, enabling the realization of low- and higher-order topological insulators (HOTIs), topological lasers, and topological entangled light sources, with promising implications for advanced photonic devices [1-5]. In these settings, translational invariance conventionally underpins momentum-space topology and the bulk–boundary correspondence that links band topology to protected boundary modes.

Recently, increasing attention has been directed to topological physics beyond periodic order, including systems with quasiperiodic or fractal geometry, where translational symmetry is partially or entirely absent. Fractal lattices have been predicted and demonstrated to host a range of unconventional topological phases, including (anomalous) Floquet photonic topological insulators [6-8], nonlinear fractal corner states [9], and HOTI phases in acoustic systems [10, 11]. Beyond boundary localization, fractal architectures have also enabled unconventional transport and localization phenomena such as internal non-Hermitian skin effects [12], fractal-induced topological phases [13, 14], and quantum transport in fractal networks [15].

At the same time, these developments highlight an intrinsic challenge of exact fractal systems: the absence of translational invariance and, in many cases, the lack of a well-defined bulk structure, precludes the direct application of conventional momentum-space topology and bulk–boundary correspondence [6, 14, 16]. Consequently, key concepts such as Brillouin zones, band gaps, and topological invariants lose their standard meaning, complicating both systematic classification and scalable design. This raises a central and unresolved question: is it possible to construct lattice geometries that retain the hierarchical richness of fractals while recovering a controllable, unit-cell-based topological description? Addressing this challenge would connect two regimes of topological matter that have so far remained largely separated, thus enabling new strategies for engineering topological states with controllable multiplicity and functionality. As a step in this direction, a class of "fractal-like" lattices has been introduced that interpolate between periodic crystals and exact fractals [17]. In these systems, different generations of a given fractal are treated as effective unit cells, restoring translational invariance while retaining self-similar internal structure. Such lattices have been shown to support multiple flat bands and real-space topological localization [18, 19]. However, how the interplay between periodicity and self-similarity governs the scaling and organization of topological boundary states remains largely unexplored.

Here we demonstrate exponential scaling of topological boundary states in photonic lattices with fractal-like geometry. As illustrated schematically in Fig. 1, both quasi-one-dimensional (quasi-1D) and two-dimensional (2D) fractal-like lattices constructed from Koch-curve and Sierpiński-gasket unit cells, respectively, support families of topological edge or corner modes whose number $N_\ell$ grows exponentially with the fractal generation index $\ell$. We develop a unified theoretical description based on our recently discovered multi-topological-phase (MTP) framework [20], which not only predicts their existence but also quantitatively determine both the exact number of boundary states $N_\ell$ and the number of topological minigaps $M_\ell$. Specifically, we find that $N_\ell = \nu M_\ell$, where $\nu$ is an integer determined by the underlying symmetry, and analyze how these quantities depend on the coupling parameters. Experimentally, we directly observe this exponential scaling of boundary mode. Experimentally, we directly observe this exponential scaling of boundary modes in laser-written fractal-like photonic lattices. The resulting boundary-state proliferation is robust, reflecting an elegant interplay between topology, fractality, and lattice periodicity. Our work establishes a general strategy for engineering fractal-structured synthetic materials and topology-enabled photonic architectures.

**Results**

**The 1D fractal-like chain lattice:**

Let us explain the scheme for constructing fractal-like lattices that combines the essence of fractality and periodicity in the first example involving Koch-curve unit cells. These lattices are constructed in generations, in a somewhat similar but in fact distinct way from the exact fractals; generations are labeled by the structural control parameter $\ell = 0,1,2,3,...$ (Fig. 2a). In the first generation $G(0)$ the unit cell has only two sites, while in the second generation $G(1)$ it has five sites corresponding to the simplest recognizable Koch-curve element. In generation $G(2)$, the unit cell has 17 sites, which already has the features of a larger Koch-curve segment. In every successive generation, the number of lattice sites (i.e., sublattices) in each unit cell increases exponentially. For the Koch fractal-like lattice of generation $G(\ell)$, the number of sublattices is $4^\ell + 1$, and the unit cells are arranged along one dimension ($x_1$) periodically, thus forming a quasi-1D periodic lattice with a fractal-like unit cell (see Fig. 2b for the first two generations). Such a scenario for constructing fractal-like lattices can be used in higher dimensions as well. The 2D fractal-like lattices constructed from Sierpiński-gasket unit cells (Figs. 2c and 2e) are discussed below.

The recently developed MTP theory [20] is highly suitable to describe the topology of fractal-like lattices. In this theory, one divides the elements of every unit cell into intra- and inter-sites, where the former do not couple to sites in other unit cells along one or more dimensions (see Methods). We apply the MTP theory to every lattice generation $G(\ell)$ separately. In essence, the $G(0)$ lattice chain is equivalent to the well-celebrated topological Su–Schrieffer–Heeger (SSH) model studied previously [21-24]. It admits two inequivalent choices of intra-sites (site A and site B, see Methods). Each sublattice defines an independent topological invariant: site A corresponds to the localized left-edge state, whereas site B corresponds to the right-edge state. An example involving A as the intra-site is illustrated in Figs. 3a-3f. (see Methods and Supplementary Information Section 1). For the $G(1)$ lattice, each Koch-curve-like unit cell contains five sublattices (A - E). Under the MTP framework, we may choose the first four sites (A-D) as intra-sites, thereby defining four distinct winding numbers $\mathcal{W}_{1-4}$ (see Supplementary Information Section 1 for details) associated with four left-edge states (Figs. 3g-3j). The winding numbers $\mathcal{W}_{1-4}$ are either 0 or 1, corresponding to the absence or presence of edge states (Figs. 3h and 3i). For a representative dimerization parameter Δ= 0.8 within the topologically nontrivial regime, the eigenvalue spectrum exhibits four topological bandgaps, each hosting two edge states (Figs. 3i and 3j).

At this stage, one might be tempted to regard these winding numbers as redundant, since they are nonzero within the same parameter regime. However, such an interpretation misses a crucial point. The MTP invariants used here not only signal the existence of boundary modes, but also quantitatively determine their number $N_\ell$. More generally, for a Koch-chain lattice of arbitrary generation $G(\ell)$, each unit cell contains $4^\ell + 1$ sublattices. The MTP theory allows the definition of $4^\ell$ independent winding numbers, which predict the emergence of $M_\ell = 4^\ell$ topological bandgaps (see Fig. 2d) and a total of $N_\ell = 2 \times 4^\ell$ edge states. Thus, $N_\ell = 2M_\ell$ where the integer ν = 2 is a consequence of inversion symmetry. **This constitutes the first explicit manifestation of the exponential sacling of topological boundary states** (Supplementary Information Sections 1 and 4).

To experimentally validate this prediction, we demonstrate the selective excitation and detection of individual topological edge modes in 1D fractal-like photonic lattices of different generations. We fabricate $G(0)$ and $G(1)$ lattices consisting of 10 unit cells and systematically probe only the left-localized edge modes in each predicted topological bandgap, given the inversion symmetry. The experimental results are displayed in Figs. 3e, 3f, 3k and 3l. The lattices are established by the continuous-wave (CW) laser-writing technique, as used in our recent work [25-30]. Inter-waveguide couplings are controlled by adjusting the waveguide spacing. The spacing is so chosen that only nearest-neighbor couplings are non-negligible, which allows us to experimentally implement the unit cell of $G(1)$ with five equidistant lattice sites in a linear arrangement, while inter-cell coupling is controlled via a zigzag geometry. To selectively excite a given topological edge state, the input probe beams are shaped in both amplitude and phase to match the calculated eigenmodes (insets of Figs. 3e and 3k), ensuring mode-selective excitation. In the topologically nontrivial lattices, the optical intensity remains localized mostly at the excited edge sites after propagation (Figs. 3e and 3k). In contrast, under identical excitation conditions, trivial lattices exhibit observable discrete diffraction (Figs. 3f and 3l). These observations directly confirm both the existence and the generation-dependent multiplicity of topological edge states, providing experimental evidence for the exponential growth predicted by theory. Numerical simulations for long-distance propagation, presented in the Supplementary Information Sections 3 and 5, further corroborate these results. The same approach can be used to verify the exponential sequence of topological bandgaps and edge states for other fractal generations.

**The 2D fractal-like lattices and higher-order topology:**

We now turn to the 2D fractal-like Sierpiński-gasket tiling lattices. For a $G(\ell)$ lattice composed of $L$ stacked layers along $x_{1,2}$ directions, the total number of unit cells is $L(L+1)/2$. Each unit cell contains $U_\ell = 3(1+3^\ell)/2$, $(\ell = 0,1,2,...)$ sublattices, yielding a total number of lattice sites $S = U_\ell L(L+1)/2$. Unlike genuine fractal lattices, which are self-similar [6-8, 10, 11, 16], the fractal-like lattices studied here retain translational invariance while embedding fractal geometry within the unit cell. This key distinction allows the definition of momentum-space topology while preserving fractal-induced hierarchy. For clarity, we focus on the first two generations $G(0)$ and $G(1)$ with an array size $L = 4$ (Fig. 2c), noting that the analysis generalizes straightforwardly to arbitrary values of $\ell$ and $L$ (Supplementary Information Sections 2 and 4).

Figure 4 summarizes the numerical and experimental results for the two generations under investigation. The $G(0)$ lattice is equivalent to the well-known breathing Kagome lattice (BKL), a paradigmatic second-order topological insulator [28, 31-37]. Its three sublattices allow three equivalent choices of intra-sites, each associated with one corner state protected by the generalized chiral symmetry and rotational symmetry [32, 38]. The real-space spectrum of the nontrivial $G(0)$ lattice exhibits a gapped structure with a set of triply degenerate zero-energy corner states (Figs. 4b–4d). Rather than reinterpreting the established BKL physics, we emphasize that $G(0)$ serves as the fundamental "seed" generation of the exponential growth mechanism within the MTP framework. As the index $\ell$ increases, the number of higher-order topological corner states grows rapidly. For the $G(1)$ lattice, there are three corner states at every band gap (Figs. 4g and 4j). Unlike the 1D case, we have three distinct winding numbers associated with three topological band gaps in the 2D tiling lattice (Figs. 4h and 4i). These invariants cannot be unified, as they are non-zero in different parameter regimes. Figure 4j highlights all HOTI corner-state eigenvalues supported by the $G(1)$ lattice, along their corresponding topological minigaps shaded in gray and labeled with $g_\ell^{(n)}$. As the number of available topological bandgaps increases, the individual gap width decreases, reminiscent of spectral minigap formation in topological quasicrystals [39]. The maximum number of topological minigaps in a $G(\ell)$ lattice is $M_\ell = 3 \times 2^{\ell-1} (\ell \geq 1)$, yielding a total number of HOTI states $N_\ell = 3M_\ell$ (in this example $\nu = 3$ is a consequence of $C_3$ symmetry), which scales exponentially with $\ell$ as shown in Fig. 2e.

To experimentally verify the exponential scaling in 2D lattices, we employ the CW-laser-writing technique [25-30] to establish photonic $G(0)$ and $G(1)$ fractal-like tiling lattices based on the Sierpiński-gaskets (as shown in Fig. 2c), in both trivial and nontrivial geometries. Probe beams are shaped in amplitude and phase to selectively excite HOTI corner modes, as shown in the insets of Figs. 4e and 4k. In the nontrivial lattices, the output intensity remains strongly confined to the excited corner sites, independent of the specific vertex chosen, reflecting $C_3$-rotational symmetry protection. In contrast, trivial lattices exhibit strong coupling to neighboring sites and delocalization (Figs. 4f and 4l). We have thus observed, in experiment, the characteristic topological corner states that are theoretically predicted to exist in the $G(0)$ and $G(1)$ lattices. These experimental observations, together with the theoretical analysis and numerical simulations (Supplementary Information Sections 2-5, provide direct confirmation of the generation-dependent exponential proliferation of higher-order topological states.

**Discussion and conclusion**

The exponential growth of topological boundary states controlled by the generation index $\ell$ represents a previously unexplored topological mechanism. The unit-cells constructed here possess fractal-inspired geometric structures; however, one can envision alternative schemes (not involving fractals) for constructing versatile lattices in generations $G(\ell)$, wherein the number of sites per unit cell also grows exponentially with $\ell$. In our 1D example of the Koch curve, because the lattice is in the tight-binding regime, we achieve the same exponential growth of topological boundary states using periodic systems with conventional unit cells (Fig. 3). Thus, we do not claim that fractality is either a necessary or sufficient ingredient for the exponential growth of topological boundary states and topological gaps. Rather, fractals serve as guiding inspiration for constructing unit cells and subsequent lattice structures, which turn out to possess intriguing characteristic of exponential growth of topological boundary states.

The fractal-like lattices constructed here are distinct from periodic lattices with simple unit cells containing a fixed number of sites [38, 40], where the number of protected boundary states is typically fixed and small; they also differ from exact fractal lattices [6, 9-11], where the absence of translational invariance precludes a conventional bulk–boundary correspondence and necessitates purely real-space topological descriptions; yet they also differ from quasicrystalline systems [39], in which the number of topological states generally depends on system size without a simple predictive analytical relation.

In the examples presented, the topological states are determined by MTP theory, which requires constrained inter-cell couplings and symmetry protection [20]. Accordingly, perturbations that respect these constraints are not expected to destroy the topological states. However, it is important to clarify that, in our construction, as the number of topological minigaps exponentially grows, the gap size decreases. Therefore, perturbations involving a certain amount of energy (no matter how small) can, in principle, affect the topological states associated with minigaps when their sizes are comparable to or smaller than the perturbation energy.

For simplicity, we fix the dimerization parameter $\Delta$ in our construction and then explore the lattice hierarchy across successive generations $G(\ell)$. As a result, in our Sierpiński-gasket example, the exponential growth of topological boundary states eventually slows down and saturates at a finite value of $\ell$. If $\Delta$ is chosen closer to unity, this saturation occurs at a larger $\ell$. By contrast, if $\Delta$ is treated as an independent degree of freedom and allowed to vary with generation, i.e., $\Delta = \Delta(\ell)$, then the exponential growth can, in principle, continue without bound. Moreover, even richer scaling behaviors can emerge, including Fibonacci sequences, sequences of positive integers, and arithmetic progressions, suggesting a broader landscape of hierarchy-driven topological phenomena that merits further exploration (Supplementary Information Section 7).

In conclusion, we have theoretically predicted and experimentally demonstrated, within the framework of multi-topological-phase (MTP) theory, a new topological scaling mechanism in periodic structures with fractal-inspired geometry, whereby the number of topological boundary states grows exponentially. Although realized here in laser-written photonic lattices, the underlying principle is general and extends to a wide range of physical platforms, including acoustic metamaterials, electrical circuits, fiber-loop systems, and synthetic frequency lattices. By exploiting hierarchical geometry, this mechanism enables an exponential increase in protected boundary modes without a commensurate expansion of bulk size or propagation length. Beyond advancing fractal-related topological physics, our results establish a conceptual link between band topology and hierarchical lattice design, suggesting a route toward compact, high-capacity topological mode architectures, with potential relevance for topological mode–division multiplexing and robust information encoding in integrated photonic systems.

## Methods

### Multi-topological-phase theory and bulk-boundary correspondence

In our fractal-like structures, the introduction of complex fractal unit cells makes it challenging to accurately describe these topological boundary states based on the conventional bulk-boundary correspondence [1, 38, 40], especially for lattices with high generation index. More specifically, it is neither feasible to precisely predict the regimes where topological states emerge nor their number. However, the topology in fractal-like systems is quite naturally described with the MTP theory [20]. As the generation index $\ell$ increases, the number of sites in each unit cell and consequently the number of topological gaps increases exponentially, while the specific intra- and inter-cell connectivity is determined, respectively, by the fractality of the unit cells and the periodicity of the whole lattice in every generation $G(\ell)$. Every topological band gap along with its associated topological state is related to its own topological invariant, given by the winding number in the MTP theory [20].

The MTP theory is designed for a lattice model with $N+J$ sublattices (i.e., sites) in a unit cell, where the first group of $N$ sublattices are the so-called intra-sites and lack inter-cell couplings along dimensions $x_i$ ($i=1$ for 1D case and $i=1,2$ for 2D case), and the second group of $J$ sublattices consists of all remaining sites. The $k$-space Bloch Hamiltonian for the presented fractal-like lattices can be written as

$$H(\boldsymbol{k}) = \begin{pmatrix} h_{\text{intra}}(\boldsymbol{k}_{\text{intra}}) & f(\boldsymbol{k}) \\ f^\dagger(\boldsymbol{k}) & \delta(\boldsymbol{k}) \end{pmatrix}, \quad (1)$$

where $h_{\text{intra}}(\boldsymbol{k}_{\text{intra}})$ is of dimension $N \times N$, and $\delta(\boldsymbol{k})$ is of dimension $J \times J$. The wavevector $\boldsymbol{k} = (k_1, k_2)^{\text{T}}$ is dual to the spatial vector $\boldsymbol{x} = (x_1, x_2)^{\text{T}}$. Within this theoretical framework, the winding numbers have the form

$$\mathcal{W}_i(\boldsymbol{k}_{\text{intra}}) = \prod_{l=n+1}^{d} \mathcal{W}_{i,l}(\boldsymbol{k}_{\text{intra}}),$$

$$\mathcal{W}_{i,l}(\boldsymbol{k}_{\text{intra}}) = \frac{1}{2\pi} \int_0^{2\pi} dk_l \frac{d\Phi_{i,l}(\boldsymbol{k}_{\text{intra}}, k_l)}{dk_l}, \quad (2)$$

where $\Phi_{i,l}(\boldsymbol{k}_{\text{intra}}, k_l)$ is the phase of $q_{i,l}(\boldsymbol{k}_{\text{intra}}, k_l) = |q_{i,l}(\boldsymbol{k}_{\text{intra}}, k_l)|e^{-i\Phi_{i,l}(\boldsymbol{k}_{\text{intra}}, k_l)}$. The Bloch Hamiltonians and the calculations of topological invariants for all models involved can be found in the Supplementary Information Sections 1 and 2.

### Laser-writing fractal-like lattices and probing topological boundary states

In experiments, the finite-sized fractal-like lattices with Koch-curve and Sierpiński-gasket unit cells were fabricated using a CW-laser-writing technique [25-29]. To optically induce a waveguide, a nonlinear bulk crystal (strontium barium niobate) was placed at the waist of the Gaussian-shaped writing beam, and the position through which the beam passes the crystal was controlled by a spatial light modulator (SLM). This procedure was repeated to create the lattices in the desired configuration with fractal Koch-curve and Sierpiński-gasket unit cells. In experiment, a zigzag-shaped lattice structure (with zigzag angle $\theta = 90°$) is used (Fig. 3) in lieu of the original Koch-curve-shaped geometry (Fig. 2b) to achieve an equivalent lattice coupling, eliminating unwanted nearest-neighbor coupling between the B and D sublattices.

To efficiently excite the topological boundary states, the SLM was used to generate the corresponding probe beams, which consist of multiple Gaussian beams with designed profiles matching the target mode distributions. The location, amplitude, and phase of each constituent Gaussian beam were precisely controlled by the SLM through the corresponding phase mask. In this way, mode-resolved excitation was achieved, suppressing crosstalk arising from different modes occupying the same lattice sites. At output from the lattice, the measured intensity and phase patterns provide the information about the presence (absence) of the edge state in nontrivial (trivial) phase, as seen also from the intensity-plot insets in Fig. 3 and Fig. 4. Additional details and simulation results corresponding to experiments are provided in Supplementary Information Sections 3, 5 and 6.

**Acknowledgments**
This research was supported by the National Key R&D Program of China (No. 2022YFA1404800); the National Natural Science Foundation of China (No. 12504384,W2541003, 12134006, 12374309, 12274242, and 124B2078); the Natural Science Foundation of Tianjin (No. 21JCYBJC00060); the Natural Science Foundation of Tianjin for Distinguished Young Scientists (No. 21JCJQJC00050); the 111 Project (No. B23045) in China; the Postdoctoral Fellowship Program of CPSF (No. GZC20252237); and the China Postdoctoral Science Foundation (No. 2025M773391). H.B. acknowledges support from the project "Implementation of cutting-edge research and its application as part of the Scientific Center of Excellence for Quantum and Complex Systems, and Representations of Lie Algebras", Grant No. PK.1.1.10.0004, co-financed by the European Union through the European Regional Development Fund— Competitiveness and Cohesion Programme 2021-2027. R.M. acknowledges support from the NSERC Discovery and the Canada Research Chair programs.

**Competing interests**
The authors declare no competing interests.

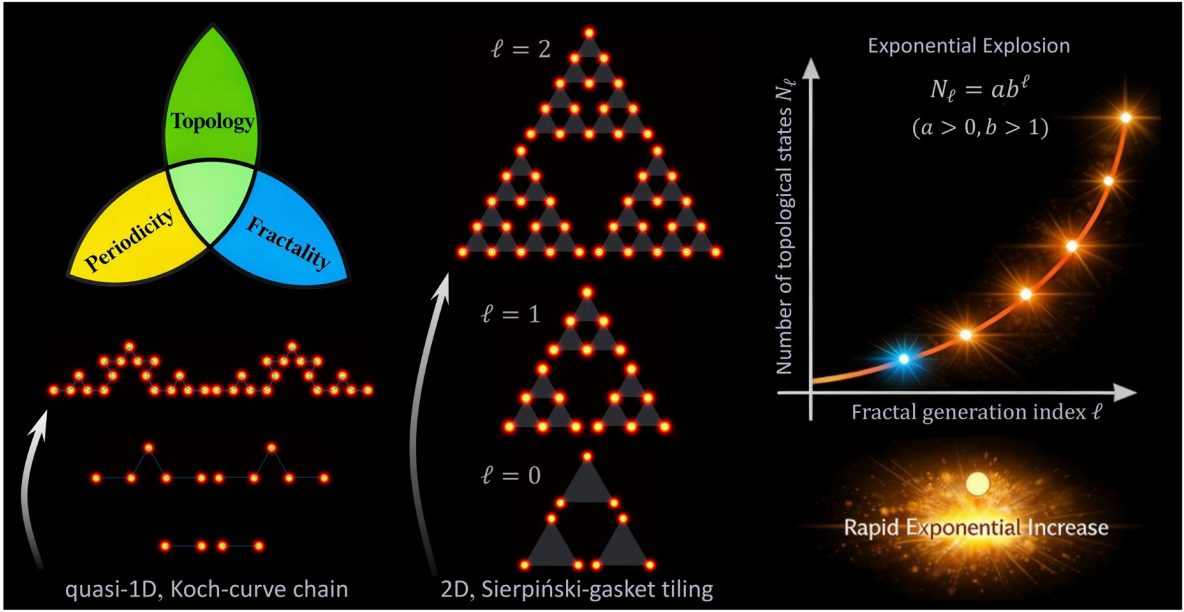

**Fig. 1 | Fractal-mediated exponential explosion of topological boundary states.** Left: Schematic illustration of the interplay between topology, fractality, and lattice periodicity (top), and recursive quasi-1D fractal-like chain lattices with Koch-curve unit cells at increasing generation index $\ell$ (bottom). Middle: 2D fractal-like tiling lattices with Sierpiński-gasket unit cells. Right: Conceptual illustration of exponential explosion of topological boundary states, showing the growth of the boundary-state number $N_\ell$ with $\ell$, following an exponential scaling $N_\ell = ab^\ell$ (where $a > 0$ denotes the symmetry coefficient, and $b > 1$ is the scale factor). This behavior highlights a fractal-enabled exponential amplification of topological boundary states beyond the linear scaling typical of conventional periodic lattices.

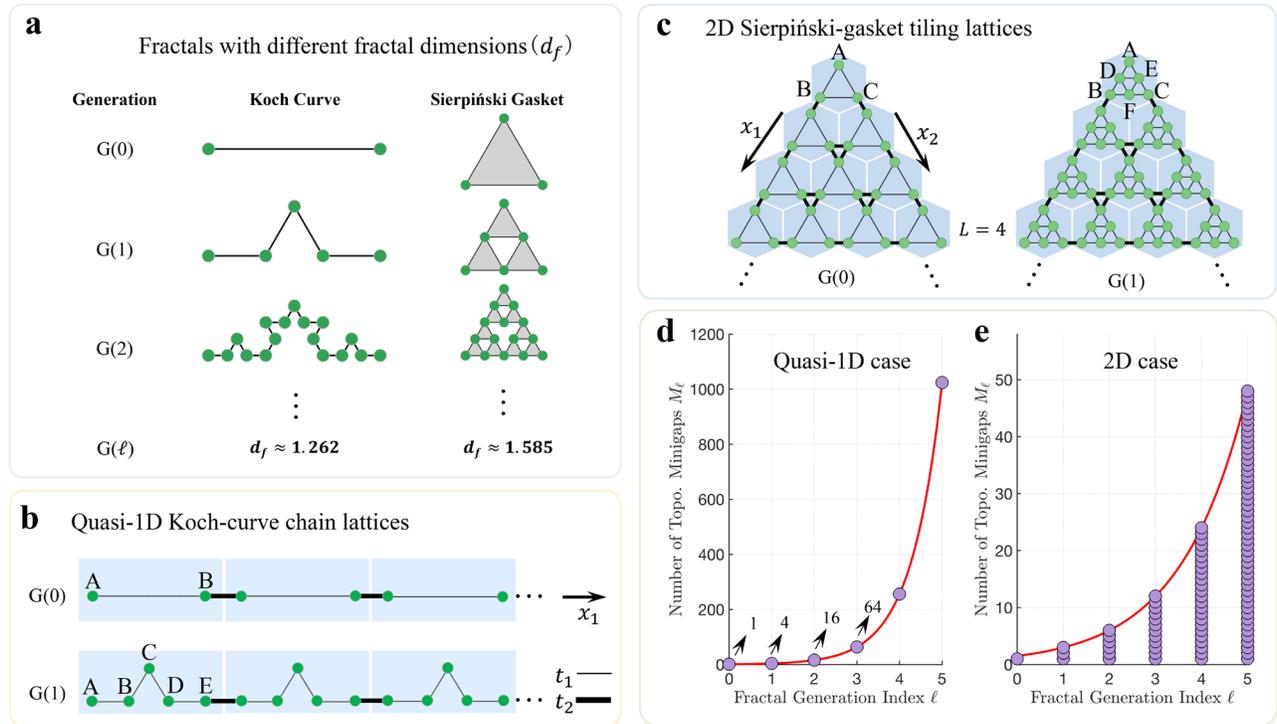

**Fig. 2 | Fractal-driven lattices and exponential sequences. a** Two representative fractal structures whose elements of generation are used to construct quasi-1D (left, Koch curve) and 2D (right, Sierpiński-gasket) periodic fractal-like lattices. **b** and **c** Schematic illustrations showing that the Koch-curve and Sierpiński-gasket unit cells in (**a**) are respectively used as unit cells to construct the first two generations of 1D and 2D fractal-like lattices by translation along $x_{1,2}$ directions. $t_1$ and $t_2$ are the intra-cell and inter-cell coupling parameters, respectively, controlled by the lattice spacings. **d** and **e** Exponential increase of the numbers of topological minigaps $M_\ell$ with fractal generation index $\ell$ for 1D (**d**) and 2D (**e**) fractal-like lattices. The total number of topological states $N_\ell$ in each generation is an integer multiple of $M_\ell$, depending on the lattice symmetry. For (**d**), each generation supports only a single topological bandgap; while for (**e**), each generation (for $\ell > 0$) can support multiple topological band gaps, i.e., $M_\ell$ can take values from 1 to max $[M_\ell]$ as the topological complexity of the lattice increases (see the example in Fig. 4i below).

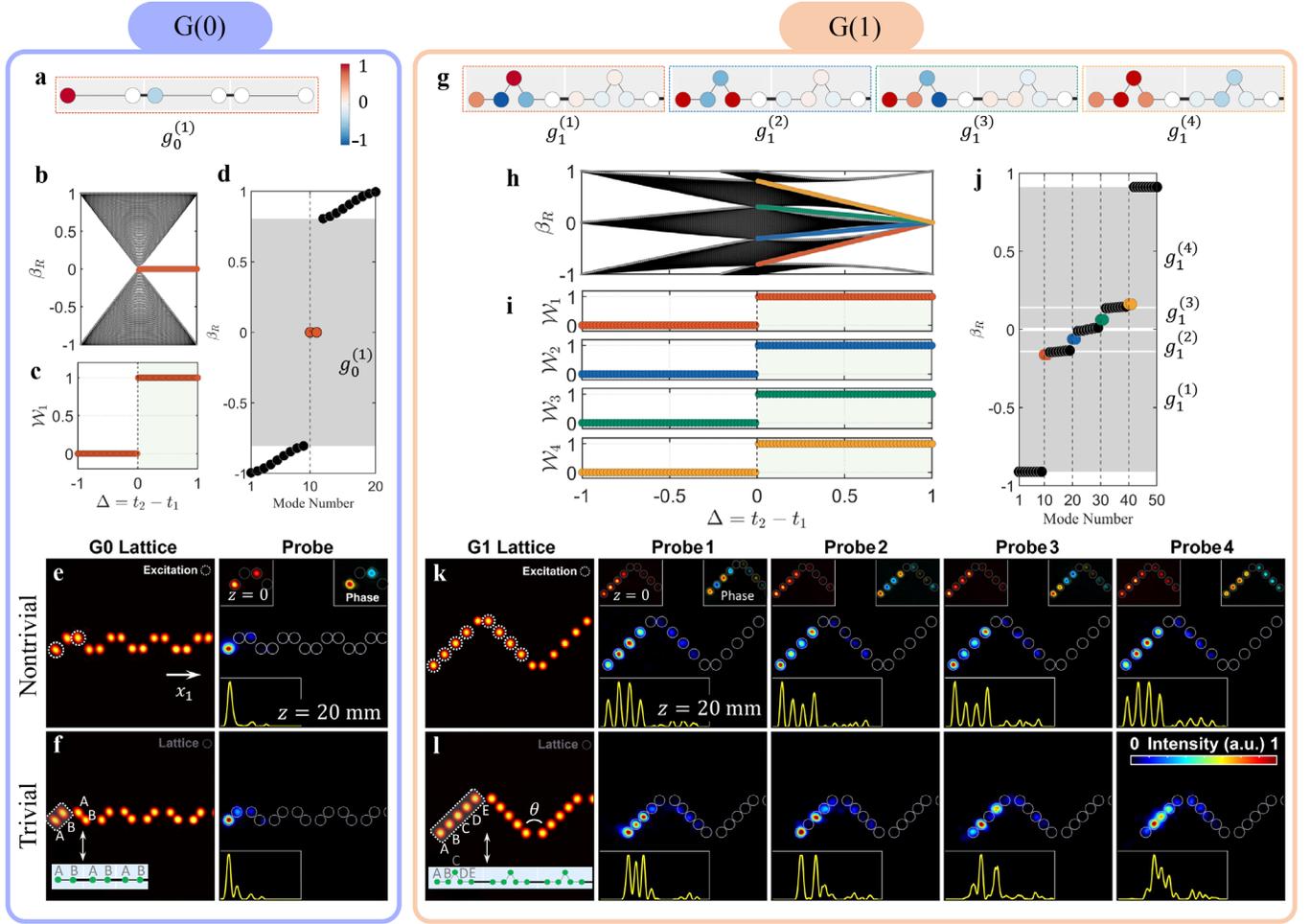

**Fig. 3 | Demonstration of exponential growth of topological edge states in quasi-1D lattices. a** Characteristic mode distribution of a representative edge state residing in the $g_0^{(1)}$ bandgap and localized at the left edge of the $G(0)$ lattice, wherein both amplitude and phase are normalized. **b** and **c** Eigenvalue spectrum $\beta_R$ (**b**) and winding number (**c**) as a function of dimerization parameter $\Delta = t_2 - t_1$. **d** Eigenvalue spectrum at $\Delta = 0.8$. The Gray region denotes the bandgap $g_0^{(1)}$ ($g_\ell^{(n)}$ refers to the location of the $n$th topological bandgap in the $\beta_R$ spectrum of a $G(\ell)$ lattice, with $n = 1, 2, \cdots, M_\ell$). **e** Experimental $G(0)$ lattice in the nontrivial phase (left panel) and corresponding edge-excitation results (right panel). The output intensity mainly occupies the A sublattice in the left edge – showing the characteristic of SSH-type edge states. The intensity and phase of the probe beam are shown in the top-left and top-right insets. **f** Same layout as that in (**e**) but for a trivial phase, where the probe beam leaks noticeably into the B sublattice under same excitation condition. **g-l** Same layout as in (**a**)-(**f**), but for the $G(1)$ lattice. In this case, four winding numbers $\mathcal{W}_{1\sim4}$ (**i**) are defined, shown together with (**j**) four topological bandgaps (gray regions, indexed by $g_1^{(1\sim4)}$) and (**g**) corresponding characteristic edge mode distributions. The intensity

profiles (colored lines, taken along each zigzag-shaped lattice) are provided to show the distinct amplitude distributions of edge states. (**k**, **l**) are corresponding experimental results.

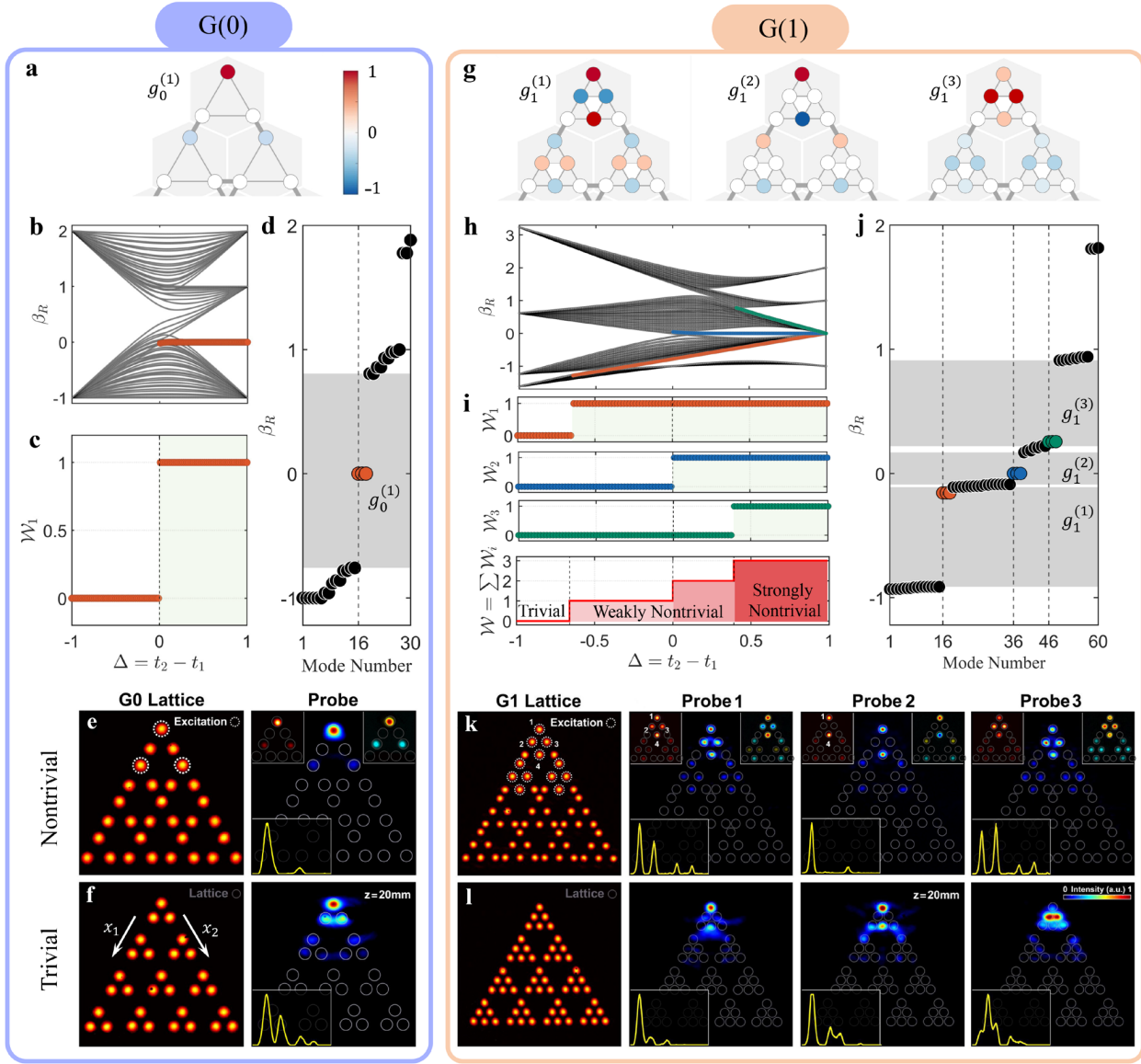

**Fig. 4 | Demonstration of exponential growth of HOTI corner states in 2D fractal-like lattices. a-l** The same layout as in Fig. 3, but now for a 2D fractal-like lattice. Nonzero corner-mode amplitudes are confined to intra-site sublattices, and the amplitude profile decays exponentially from the lattice vertex into the bulk, exhibiting either out-of-phase or in-phase relations among neighboring layers of unit cells. Here, three winding numbers $\mathcal{W}_{1\sim3}$ are defined, and three topological bandgaps (gray regions, indexed by $g_1^{(1\sim3)}$) and corresponding characteristic corner mode distributions are given. The total winding number $\mathcal{W} = \sum \mathcal{W}_i$ determines the number of topological bandgaps under weakly and strongly nontrivial regimes, which are separated by the dimerization parameter $\Delta$. The intensity profiles are taken along $x_2$ direction

at the lattice right boundary. Additional simulated results and robustness analyses are provided in the Supplementary Information Sections 3, 5 and 6.